# ChemFlow：A Hierarchical Neural Network for Multiscale Representation Learning in Chemical Mixtures

Jinming Fan [a b c], Chao Qian [a c], Wilhelm T. S. Huck [b], William E. Robinson [b], Shaodong Zhou [a c*]

[a] *College of Chemical and Biological Engineering, Zhejiang Provincial Key Laboratory of Advanced Chemical Engineering Manufacture Technology, Zhejiang University, 310027 Hangzhou (P. R. China)*

[b] *Institute for Molecules and Materials, Radboud University, Nijmegen, The Netherlands*

[c] *Institute of Zhejiang University – Quzhou, 324000 Quzhou (P.R. China)*

*E-mail:* [szhou@zju.edu.cn](mailto:szhou@zju.edu.cn)

***Abstract:*** *Accurate prediction of the physicochemical properties of mixtures of molecules using graph neural networks remains a significant challenge, as it requires simultaneous embedding of intramolecular interactions as well as accounting for composition of the mixture (i.e. concentrations and ratios). Existing approaches are ill-equipped to emulate realistic mixture environments, where densely coupled interactions propagate across hierarchical levels—from atoms and functional groups to entire molecules—and where cross-level information exchange is continuously modulated by composition. To bridge the gap between isolated molecules and realistic chemical environments, we present ChemFlow, a novel hierarchical framework that integrates atomic, functional group, and molecular-level features, facilitating the flow of information across these levels to predict the behaviors of complex chemical mixtures. ChemFlow employs an atomic-level feature fusion module, Chem-embed, to generate context-aware atomic representations influenced by the mixture's state and atomic characteristics. Next, bidirectional group-to-molecule and molecule-to-group attention mechanisms allow ChemFlow to capture the interactions of functional groups both within and across molecules in the mixture. By dynamically adjusting representations based on concentration and composition, ChemFlow excels at predicting concentration-dependent properties and significantly outperforms state-of-the-art models in both concentration-sensitive and non-concentration-dependent systems. Extensive experiments validate ChemFlow's superior accuracy and efficiency in modeling complex chemical mixtures.*

## Introduction

Data-driven machine learning (ML) techniques, [1-6] especially deep learning, have emerged as a powerful paradigm to accelerate material discovery and chemical process design.[7-10] Through complex modeling of large datasets, ML models,[11-18] and especially Graph Neural Networks (GNNs),[19,20] can be trained to correlate the structural features of molecules with their chemical properties. GNNs treat molecules as graphs, classifying atoms and bonds as nodes and edges, respectively, thus enabling efficient information collection and passing.[21-22] However, this approach primarily captures intramolecular interactions and relies on non-linear transformations of atomic features and limited chemical descriptors (e.g., hybridization, valence).[23] To achieve a comprehensive prediction of molecules in a complex chemical environment, the model must be augmented with proper descriptors of both intrinsic atomic properties and the compositional interplay.[24] However, the development of descriptors that comprehensively capture the intrinsic nature of intermolecular interactions and simultaneously incorporate the concentration-dependent effects on these interactions remains an open and challenging problem.

The conventional graph-based molecular representations used by GNNs, while effective for single molecules,[20-21] are inherently limited to intramolecular interactions. This makes them unsuitable for modeling multi-molecular systems, where intermolecular interactions, solvent-induced perturbations, and composition-dependent behaviors drive the emergent physicochemical properties. Recent efforts to extend GNNs to solution-phase systems have sought to address this challenge by introducing dual-network architectures to separately encode solutes and solvents.[25-26] These approaches allow for cross-molecular message passing, while methods like 3D structure-based pretraining improve the model's robustness to conformational variability.[27] However, these models remain heavily pairwise, molecule-specific, or tightly coupled to specific architectures, making them difficult to generalize to complex, ternary or higher-order mixtures, incorporate concentration effects, or aggregate data from heterogeneous molecular components.

Furthermore, some models treat molecules as nodes and attempt to learn macroscopic interactions at the molecular scale.[28-29] While effective for coarse correlations, such abstractions struggle to capture the chemical origin of interactions in mixtures, which is often governed by functional-group interactions and group–molecule couplings. In particular, they overlook that atomic descriptors should change with chemical context, that functional groups can communicate across molecular boundaries, and that composition can modulate interaction propensity. A coherent framework is therefore needed to bridge atomic chemistry with mesoscale organization in mixtures.

To address this gap, we propose ChemFlow, as illustrated in Fig. 1a. At the atomic scale, Chem-Embed performs mixture-aware multimodal feature fusion by integrating diverse chemical descriptors from multiple sources. Specifically, atomic- and element-level features are fused through a hierarchical multi-stage process, enabling complementary information to be incorporated across representation levels. The hierarchical modulation process adjusts each atom's features at different levels, ensuring that the atom-level features adapt to the chemical environment while integrating the mixture effects. This fusion of diverse features forms a strong foundation for downstream interaction modeling. Building upon atomic representations, features are progressively aggregated to the group and molecular levels. A bidirectional attention mechanism between group and molecular representations is employed to model potential intra- and intermolecular interactions within mixtures. As shown in Fig. 1b, a concentration-aware control module further modulates representations across atomic, group, and molecular levels, enabling the model to account for variations in mixture composition and their influence on interaction patterns. Collectively, these design choices restore continuity from atomistic descriptors to mesoscale organization and ultimately to mixture-level properties, providing a unified and chemically faithful representation of complex mixtures.

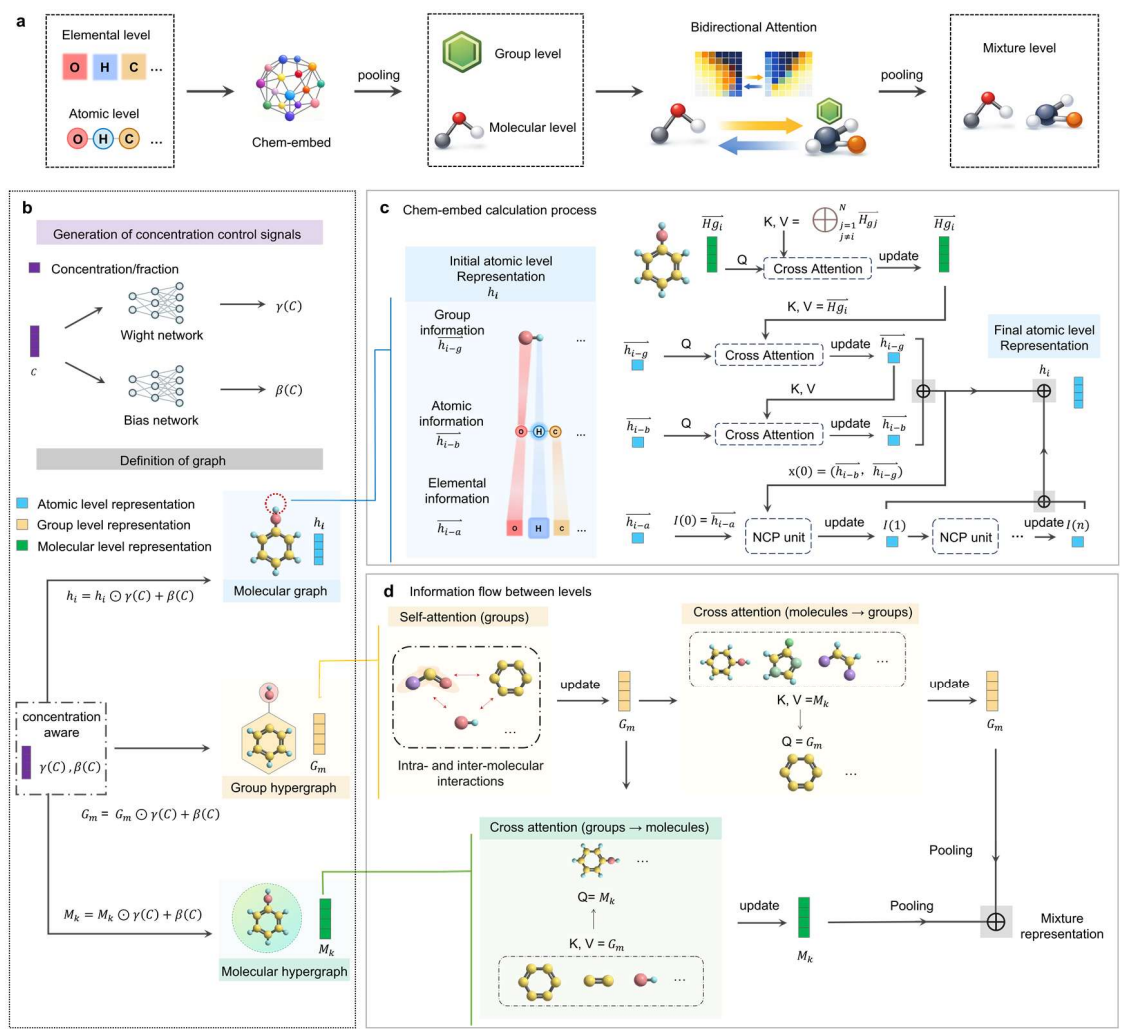

**Figure 1.** Schematic diagram of of ChemFlow. (a) Conceptual diagram of the model. (b) Definition of Molecular graph, Group hypergraph, and Molecular hypergraph, and the mechanism of concentration aware modulation on diverse graphfeatures. (c) Chem-Embed for mixture-aware multimodal atomic fusion. (d) Bidirectional group-molecule interactions in ChemFlow.

## Overview of ChemFlow

### Graph Definitions

**Molecular Graph.** For a mixture sample $s$, we consolidate atoms from all $K_s$ component molecules and concatenate them in a fixed order. Let $\mathcal{V}^{(s)} = \{v_1, \ldots, v_{N_s}\}$ be the $N_s$ atoms with initial atomic features $h_i$. The molecular graph is

$$\mathcal{G}_a^{(s)} = \left(\mathcal{V}^{(s)}, \mathcal{E}_{atom}^{(s)}\right),$$

where $\mathcal{E}_{atom}^{(s)}$ represents the set of intramolecular chemical bonds. For any bond $(i,j) \in \mathcal{E}_{atom}^{(s)}$, the edge feature is denoted as $e_{ij}$.

*Functional-group instances (hyperedges) and incidence.* We predefine a functional-group vocabulary (customizable per domain knowledge) and detect group instances by substructure matching. Let $G_s$ be the number of matched instances. For the $m$-th group instance ($m \in \{1, \ldots, G_s\}$), let $A_m \subseteq \{1, \ldots, N_s\}$ be its member-atom index set, defining a hyperedge:

$$e_m^{group} = \{v_i \mid i \in A_m\}$$

Equivalently, the atom–group relation is encoded by an incidence matrix $\mathbf{B}^{(s)} \in \{0,1\}^{N_s \times G_s}$ with

$$B_{i,m}^{(s)} = 1 \iff i \in A_m$$

stored in sparse COO format (equivalently, a star-expansion bipartite incidence graph).

*Molecular membership and incidence.* Each atom belongs to exactly one component molecule. Let $mol(i) \in \{1, \ldots, K_s\}$ denote the component index of atom $i$. The $k$-th molecule corresponds to the atom set

$$M_k = \{i \mid mol(i) = k\}$$

and can be equivalently represented by $\mathbf{S}^{(s)} \in \{0,1\}^{N_s \times K_s}$ with

$$S_{i,k}^{(s)} = 1 \iff i \in M_k$$

implemented as bool masks in practice.

*Composition signal and concentration-awareness.* Let $c$ denote the mixture composition/concentration encoding (scalar or vector), used to modulate representations via learnable functions $\gamma(c)$ and $\beta(c)$. As illustrated in Fig. 1b, this modulation mechanism assigns composition-dependent weights and biases to representations at the atomic, group, and molecular scales, thereby adjusting the strength of feature interactions under different mixture compositions. This design enables the model to adapt its internal representations according to changes in mixture composition, allowing different interaction patterns to emerge across hierarchical levels.

**Computational Pipeline**

**Chem-Embed: Mixture-aware Multimodal Atomic Representation.** As illustrated in Fig. 1c, Chem-Embed produces mixture-conditioned atomic embeddings via multimodal, feature-level fusion under a hierarchical conditioning pathway. For each atom $v_i$, we organize context signals from coarse to fine as: (i) global descriptors of the other components ($\bigoplus_{\substack{j=1 \\ j \neq i}}^{N} \overrightarrow{H_{gj}}$) in the mixture together with their concentrations ($c$), (ii) the global descriptor of the atom's parent molecule ($\overrightarrow{H_{gi}}$) together with its concentration ($c$), (iii) the atom's functional-group assignment encoded as a one-hot vector ($h_{i-g}$), and (iv) local environment descriptors of the atom ($h_{i-b}$, e.g., valence state and hybridization). This hierarchy reflects a core premise of mixtures: atomic behavior is not solely determined by intrinsic structure, but is dependent on the surrounding components and composition (for more details on descriptors, see Method). Formally, chem-Embed injects mixture effects sequentially along the hierarchy:

$$(\bigoplus_{\substack{j=1 \\ j \neq i}}^{N} \overrightarrow{H_{gj}}, c) \Rightarrow (\overrightarrow{H_{gi}}, c) \Rightarrow h_{i-g} \Rightarrow h_{i-b}$$

implemented with cross-attention so that higher-level signals modulate the lower-level descriptors. Concretely, the "other-components" context provides a global mixture state that conditions the parent-molecule representation; the parent-molecule context then regulates which functional-group cues are emphasized for the atom; finally, the group-conditioned signal refines the atom's local environment descriptors, yielding an atomic representation that is both chemically expressive and composition-aware.

After hierarchical conditioning, we collect the environment-dependent features as the initial hidden state

$$x(0) = cat(h_{i-g}, h_{i-b}, c)$$

The elemental feature vector $I(0) = h_{i-a}$ is then fused with $x(0)$ in a (Closed-Form Continuous-Time) CFC based (Neural Circuit Policies) NCP unit.[30-31] This fusion process allows the model to refine atomic representations and extract atomic features

$h_i$.

**Hierarchical Representation and information flow between levels**

After obtaining the context-conditioned atomic embeddings through Chem-Embed, ChemFlow proceeds to construct higher-level representations through hierarchical aggregation. The atomic features $h_i$ are first aggregated to form functional-group-level embeddings by pooling over each group's constituent atoms. Specifically, for each group the features of its constituent atoms $\{h_i \mid i \in A_m\}$ are pooled to the group embedding $G_m$:

$$G_m = pooling\{h_i \mid i \in A_m\}$$

where $A_m$ is the index set of atoms in the $m$-th group instances, these group embeddings $G_m$ represent the chemically significant motifs and are enriched by incorporating the environmental context provided by atomic-level features.

Next, we apply concentration-dependent modulation to adjust the group embeddings according to the mixture's composition. This is achieved through: $G_m = G_m \odot \gamma(c) + \beta(c)$, where $\gamma(c)$ and $\beta(c)$ are learnable parameters modulated by the concentration $c$, allowing the group representations to be sensitive to the changing composition of the mixture.

To compensate for potential loss of atomic details and key inter-atomic interactions during the atom-to-group aggregation, we perform message passing between atoms within each molecule to capture atomic-level interactions. This message passing process allows the atom-level features to exchange information, which helps preserve essential atomic details (the specific computation details of the message passing mechanism can be found in the Method section).

After the message passing step, we further aggregate the updated atomic representations within each molecule to obtain the molecule-level embeddings. This aggregation serves to enrich the group-level representation by recovering the atomic-level details that might be lost during the atom-to-group pooling process. The molecule-level embedding $M_k$ is obtained by pooling over the atomic embeddings $\{h_i \mid i \in M_k\}$ for each molecule:

$$M_k = pooling\{h_i \mid i \in M_k\}$$

followed by concentration-dependent modulation: $M_k = M_k \odot \gamma(c) + \beta(c)$. By modulating atomic, group, and molecule-level features with concentration, ChemFlow ensures that the features at each hierarchical level reflect their respective proportion in the mixture.

**Self-Attention: Intra- and Inter-Molecular Group Interactions.** Once the group embeddings are computed, we apply self-attention over the entire set of functional-group embeddings $\{G_m\}$ to capture intra- and inter-molecular interactions. This self-attention mechanism allows functional groups to exchange information both within the same molecule and across molecules in the mixture, facilitating the discovery of cooperative interactions. Formally, we compute the updated group embeddings $G_m$ as follows:

$$G_m = \text{SelfAttention}(\{G_m\}_{m=1}^{G_s})$$

This step captures the mesoscale interactions between functional groups, allowing the model to learn how local substructures (i.e., functional groups) influence each other within the chemical system.

**Cross-Attention: Group–Molecule Interactions.** At this stage, the model introduces cross-attention between functional groups and component molecules (Fig. 2b). This bidirectional interaction ensures that both group-level and molecule-level representations influence each other. Specifically, each group embedding $G_m$ attends to all molecule-level representations $M_k$, and vice versa, to capture global mixture dynamics.

*Groups to molecules:* Groups influence molecules by modulating the component-level representations based on the group-specific interaction propensity. Formally:

$$M_k = \text{CrossAttention}(Q = M_k, K = G_m, V = G_m)$$

where $Q = M_k$ is the query (molecule-level embedding), $K, V = G_m$ are the key and value (group-level embeddings), this mechanism updates the molecule-level embeddings $M_k$, incorporating information from the functional groups.

*Molecules to groups*: Similarly, molecule representations $M_k$ attend to all group embeddings $G_m$ to provide context and adjust group representations:

$$G_m = \text{CrossAttention}(Q = G_m, K = M_k, V = M_k)$$

which updates the group-level embeddings $G_m$, refining group-level features based on the molecular environment.

Finally, the updated molecule-level and group-level embeddings are pooled to generate a global mixture-level representation.

## Results and Discussion

**Dataset**

To comprehensively evaluate the capacity of ChemFlow to model complex mixture systems, we curated a set of representative datasets from the literature, each reflecting a distinct class of physicochemical behavior in multi-component environments.

**Concentration-dependent multi-component datasets**: 1. Activity coefficients[29]: Binary and ternary mixture datasets covering a broad range of composition ratios (mole fractions) and capturing how non-ideal intermolecular interactions evolve with mixture composition. 2. MixSolDB[28]: Solubility measurements of diverse solutes in single-, binary-, and ternary-solvent systems spanning a wide range of solvent mixing ratios, reflecting the sensitivity of dissolution processes to multi-component solvent environments. 3. Surface tension[32]: Single- and binary-solute systems with varying concentrations, providing a benchmark for modeling interfacial behavior that depends critically on solute abundance and cooperative molecular interactions.

All concentration-dependent datasets were reformulated into a unified mixture-modeling task (for example, merging datasets on binary and ternary mixtures by filling entries with a fraction of 0), enabling ChemFlow to learn from heterogeneous chemical systems within a single representational framework.

**Non–concentration-dependent photophysical dataset**: To assess the generality of the model beyond composition-varying systems, we additionally included properties of in single solute–solvent pairs[33], including: 1. absorption wavelength[33]. 2. emission wavelength[33]. 3. life time[33]. 4. Solubility: CombiSolv[34] dataset.

**Table 1**. Prediction results of concentration-dependent multi-component and binary solvent–solute dataset using different models.

| Dateset | Concentration-Dependent Multi-Component | | | | | Binary Solvent–Solute | | | |
|---|---|---|---|---|---|---|---|---|---|
| | Activity coefficient | | surface tension | MixSolDB | | Absorption wavelength | Emission wavelength | Life time | CombiSolv |
| modeling approach | Ternary | Combined binary and ternary | Combined single and binary | Binary | Combined single, binary and ternary | Binary | Binary | Binary | Binary |
| Dataset size | 160000 | 360000 | 892 | 29158 | 56789 | 17276 | 18141 | 6961 | 8780 |
| AttentiveFP | N/A | N/A | N/A | N/A | N/A | 22.86 | 28.70 | 0.871 | 0.471 |
| CIGIN[25] | N/A | N/A | N/A | N/A | N/A | 19.66 | 25.84 | 0.821 | 0.464 |
| CGIB[26] | N/A | N/A | N/A | N/A | N/A | 18.37 | 24.52 | 0.808 | 0.448 |
| CGIB+3DMRL[27] | N/A | N/A | N/A | N/A | N/A | 17.93 | 23.92 | 0.733 | 0.442 |
| SolvGNN[29]/ Subgraph GNN[28] | 0.0370 | 0.0294 | 2.2259 | 0.67 | 1.07 | 22.00 | 26.34 | 0.789 | 0.523 |
| NGNN[24] | 0.0380 | 0.0305 | 2.0255 | 0.45 | 0.87 | 20.22 | 26.55 | 0.837 | 0.482 |
| ChemFlow | 0.0205 | 0.0156 | 1.4391 | 0.088 | 0.57 | 17.24 | 23.34 | 0.722 | 0.403 |

Note: "N/A" indicates that the model does not support concentration-dependent or multi-component predictions. All training was conducted using a five-fold cross validation with a training: validation: test of 0.8:0.1:0.1. The evaluation criteria for Concentration-Dependent Multi-Component are MAE. The evaluation criteria for Binary Solvent–Solute are RMSE.

**Benchmark Comparison**

We used data sourced from the literature, to better compare the performance of our model, we compared the results with those from the best-performing models (CIGIN[25], CGIB[26], CGIB+ 3DMRL[27], SolvGNN[29], SubgraphGNN[28], NGNN[24]) in the literature. All training was conducted using a five-fold cross validation with a training: validation: test of 0.8:0.1:0.1, and the error was the average of the test set results obtained from the best five training validation sets. The concentration dependent dataset results are the average of a single five-fold cross validation, which is the average of five training iterations. The non-concentration dependent dataset is the average of 3 different five-fold cross validations, i.e. the average of 15 training iterations). The evaluation index for concentration dependent multi-component datasets is MAE, and the evaluation index for non-concentration dependent properties is RMSE.

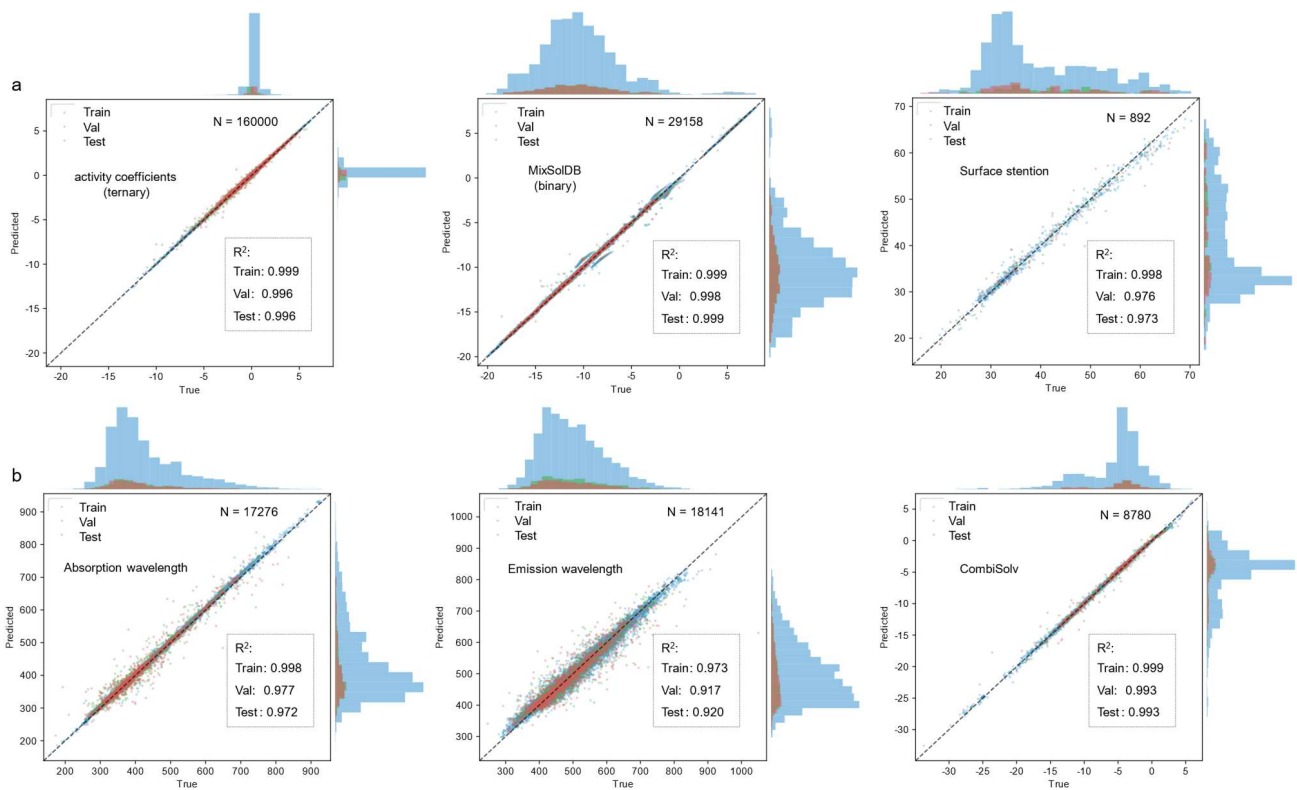

**Figure 2.** Parity plots for different datasets, with marginal histograms showing data distributions. (a) Results for the concentration dependent dataset. (b) Results for the non-concentration dependent dataset. Note: Blue, green, and red points denote training, validation, and test sets, respectively. Marginal histograms show the distributions of the true and predicted values.

We compared ChemFlow against state-of-the-art mixture prediction models across all datasets (Table 1). ChemFlow achieves SOTA performance on both non–concentration-dependent multi-component datasets and single solute–solvent tasks. Importantly, the advantage becomes especially pronounced in the concentration-dependent multi-component datasets (Activity coefficient, surface tension and MixSolDB), where intermolecular interactions are stronger, highly nonlinear, and more challenging to model. As shown in Figure 2a, it maintains high accuracy in predicting activity coefficient, solubility (MixSolDB), and surface tension, and its advantage is particularly significant in a large number of datasets (test set: $R^2>0.996$).

Notably, ChemFlow surpasses even models that incorporate 3D structure–based pretraining (CGIB+ 3DMRL[27]) on single solute–solvent tasks (Absorption wavelength, Emission wavelength, Life time and CombiSolv) —despite ChemFlow using no pretraining in our experiments. This highlights the generality of our multiscale representation: ChemFlow captures fundamental chemical principles that remain consistent across systems, not relying on dataset-specific architectural tailoring or extensive pretraining for each solute–solvent pair. As further illustrated in Table 1 and Figure 2b, on single solute–solvent tasks, ChemFlow shows more substantial improvements on datasets associated with stronger intermolecular interactions (CombiSolv), while maintaining high predictive accuracy (Figure 2b). In contrast, for datasets in which intermolecular interactions play a comparatively less dominant role (e.g., absorption and emission wavelengths), the performance gain over competing models is more modest, and the overall accuracy is relatively lower (Figure 2b). This trend is consistent with the interpretation that ChemFlow is particularly effective at modeling properties governed by intermolecular interactions.

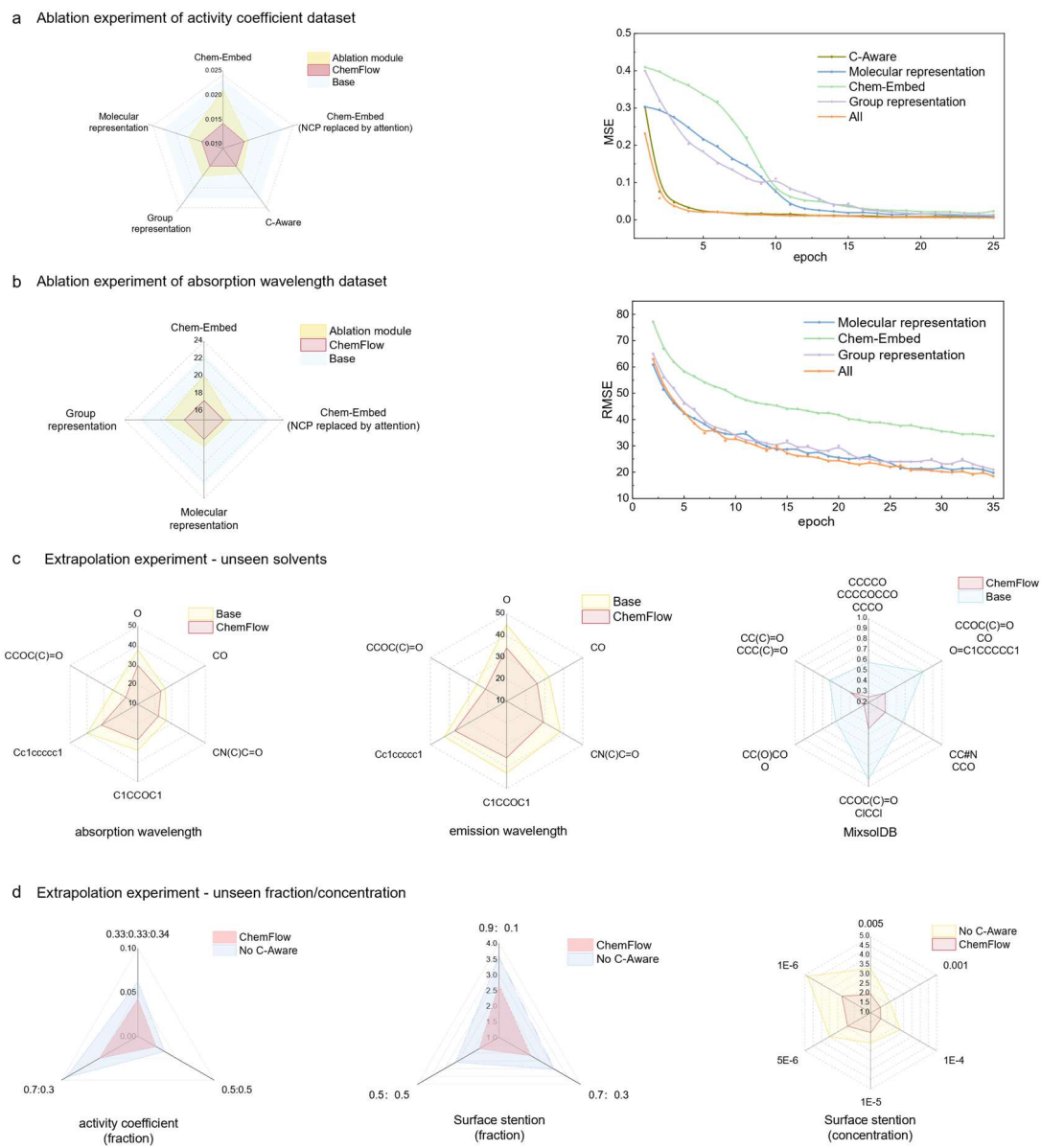

**Figure 3.** Ablation experiments and extrapolation experiments. (a) Ablation results on the concentration-dependent multi-component dataset (activity coefficients) showing the impact of removing different modules on prediction error (MAE), and the impact of removing different modules on convergence speed. (b) Ablation results on the non-concentration-dependent binary dataset (absorption wavelength) showing the impact of removing different modules on prediction error (RMSE), and the impact of removing different modules on convergence speed. (c) Extrapolation performance (RMSE) for solvents in absorption wavelength, emission wavelength, and solubility datasets. (d) Extrapolation performance for unseen fractions and concentrations with and without the concentration-aware module (C-aware).

**Ablation and Extrapolation Experiments.** To validate the contribution of each model component and assess the predictive capability in unseen domains, we conducted systematic ablation and extrapolation experiments. As shown in Figures 3a and 3b, we selected the concentration-dependent activity coefficient dataset and the non–concentration-dependent absorption wavelength dataset to perform ablation studies.

These experiments investigate the impact of removing individual modules on model performance. For the Chem-Embed module, two ablation strategies were considered: (i) replacing it with a linear layer, and (ii) substituting the Neural Circuit Policies (NCP) module with an attention layer to compute atomic interactions and update atomic features. For the concentration-aware (C-aware) module, we removed all concentration-dependent modulations at the atomic, group, and molecular levels. To analyze the role of inter-group and inter-molecular attention mechanisms, we further conducted ablations by removing group-level and molecule-level representations, respectively. In addition, we included a base model, in which all specialized modules were removed and the prediction was made solely by aggregating message-passed molecular graph representations into a mixture representation.

Figure 3 demonstrates that all modules contribute considerably to overall performance, with Chem-Embed playing a particularly critical role. Removing this module leads to a substantial degradation in prediction accuracy, indicating that Chem-embed serves as a foundational component of the model by providing meaningful atom-level representations that facilitate subsequent feature interactions. Furthermore, we analyzed the effect of different modules on training efficiency. The presence of Chem-Embed significantly accelerates model convergence, likely because it supplies each atom with a prior estimate of its behavior in the mixture, enabling faster and more efficient parameter updates during training.

For extrapolation experiments, we first evaluated the model's performance on unseen solvents, as shown in Figure 3c. The model exhibits a pronounced advantage over the base model in solubility prediction, while the improvements for absorption and emission wavelengths are less significant. This observation may be attributed to the fact that solvent variation has a relatively minor influence on spectral wavelengths but a much stronger effect on solubility, which is consistent with chemical intuition.

Furthermore, we conducted extrapolation experiments on unseen mixture fractions and concentrations in the activity coefficient and surface tension datasets, comparing the full model with a variant that excludes the concentration-aware module (Figure 3d). The results demonstrate that the C-aware module substantially enhances the model's extrapolation capability, highlights the importance of the C-aware module, which effectively modulates the representation of features in response to variations in concentration and composition.

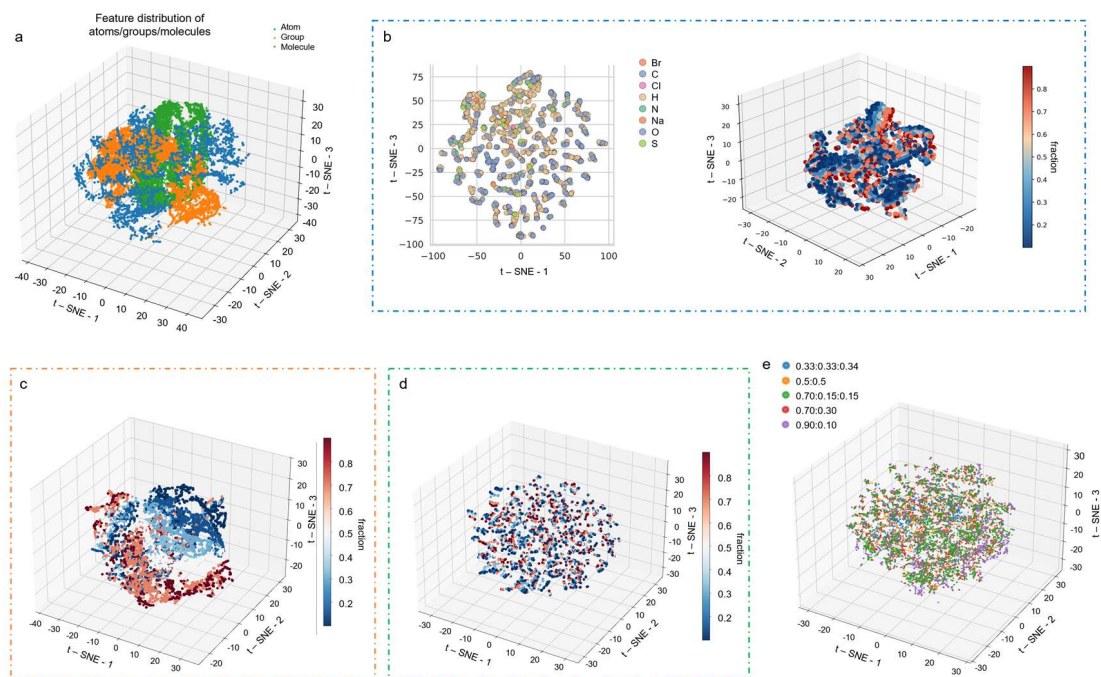

**Figure 4.** Visualization of different hierarchical features (atomic-level, group-level, molecular-level and mixture-level) in the activity coefficient dataset. (a) 3D t-SNE distribution of atomic, group, and molecular features in all mixture data within the same training batch. (b) The distribution of atomic features in mixture by the type and fraction. (c) The distribution of group-level representation in mixture by fraction. (d) The distribution of molecular-level representation in mixture by fraction. (e) The distribution

of mixture-level representation by fraction, different colors represent mixtures with different composition fractions.

**Hierarchical Feature Representations and Concentration Dependence in mixture.** To further probe what the model has learned, we perform a feature-level visualization study on the activity coefficient dataset. This dataset is particularly suitable because it is large-scale and covers hundreds of chemically diverse molecules across a wide range of mixture fractions, enabling a fine-grained examination of both composition- and concentration-driven effects.

As shown in Fig. 4a, the 3D t-SNE projection reveals that the model learns hierarchically structured representations at three levels—atomic, group, and molecular—that occupy distinct yet partially overlapping regions in embedding space. This organization indicates that the network does not collapse all information into a single latent space. In 3D space, the group-level embeddings appear to act as an intermediate bridge between atom-level local chemistry and molecule-level global context, consistent with the intended design of hierarchical message passing and aggregation.

At the atomic level, the distribution colored by atom type (Fig. 4b, left) shows that atomic embeddings exhibit clear chemical differentiation, where atoms with the same elemental identities form separable sub-structures. Importantly, when coloring the same atomic embeddings by mixture fraction (Fig. 4b, right), the points display a locally smooth, continuous transition rather than abrupt switches—atoms occupy nearby regions while their colors gradually change. This suggests that the model encodes not only intrinsic atomic identity but also how the atomic environment is modulated by concentration, capturing fraction-dependent local interactions.

Moving to higher levels, the group-level representations (Fig. 4c) show an even more pronounced fraction-dependent organization: the embedding manifold exhibits a clear gradient where nearby points correspond to similar fractions, and the fraction varies smoothly along the geometry of the latent space. This implies that functional groups serve as a sensitive carrier of fraction effects—groups aggregate atom-level signals into mesoscopic descriptors that strongly reflect composition-controlled interaction patterns, such as preferential association or repulsion between specific motifs in mixtures.

Next, the molecular-level embeddings (Fig. 4d) maintain a more globally mixed distribution while still preserving local fraction continuity. Compared with group-level features, the molecular space appears less strictly ordered by fraction, which is expected: molecule-level representations must simultaneously encode global structure and identity across many different species, so fraction becomes one (important) factor among multiple global determinants. Nevertheless, the absence of sharp boundaries and the presence of local transitions indicate that the model captures continuous concentration-driven shifts at the molecule level as well, rather than treating different fractions as unrelated regimes.

At the mixture level (Fig. 4e), embeddings appear dispersed and irregular, with no specific order tied to fractions. This suggests that the model does not impose a rigid compositional hierarchy but instead synthesizes atomic, group, and molecular features to navigate chemical mixtures, integrating multi-level information to handle the complexity of chemical systems.

In summary, across all analyzed levels, the model demonstrates its ability to connect and integrate information, effectively handling the multi-scale nature of chemical systems. Importantly, as the level of complexity increases (atomic → group → molecular → mixture), the dependence of feature distributions on mixture fraction gradually decreases. This occurs because at the atomic and group levels, local chemical environments and functional group interactions are highly sensitive to mixture composition, making fraction changes significantly influence embedding distributions. In contrast, at the molecular and mixture levels, representations must integrate more global structural information and complex inter-species interactions, so the influence of a single fraction factor is relatively diluted, resulting in reduced fraction dependence.

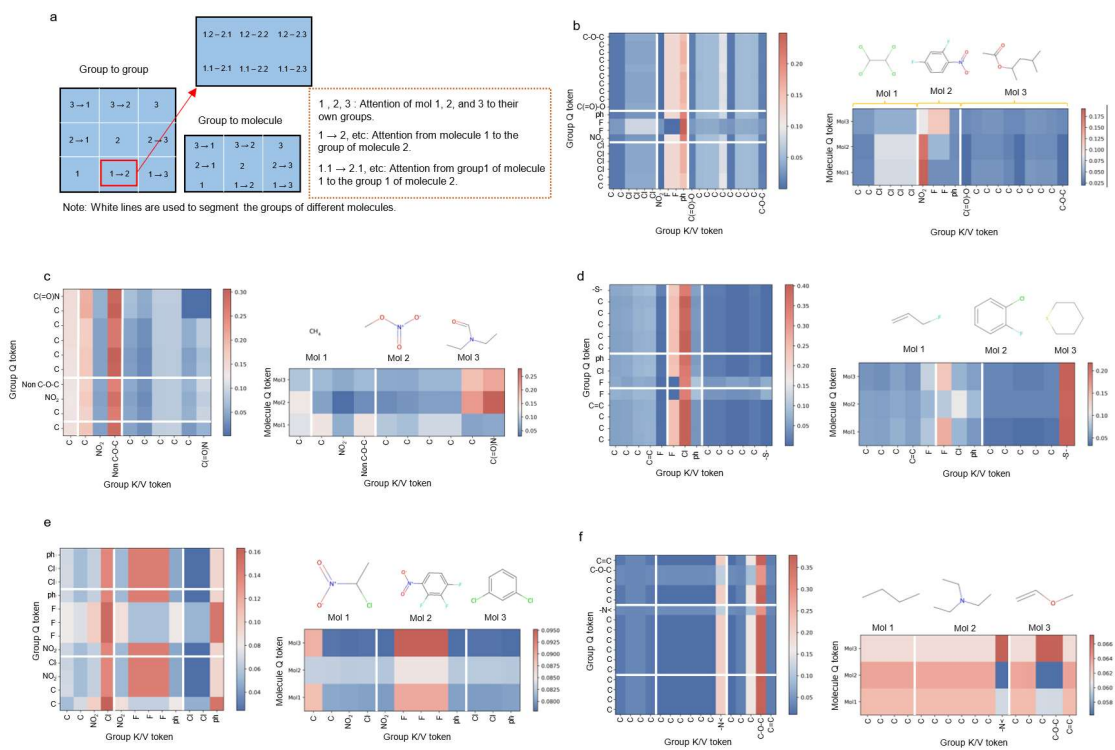

**Figure 5.** Visualization of attention maps for different mixtures in the activity coefficient dataset (a) Introduction to the content represented by each region of the attention matrix. (b) Attention maps for five different mixtures.

**Attention Mechanism Analysis at the Group–Molecule Interface.** To gain deeper insight into how the model captures interactions both between functional groups and between groups and molecules within mixtures, we selected several mixtures involving smaller molecules. We then visualized the learned attention maps for five mixtures in the activity coefficient dataset, as shown in Fig. 5. For clarity, Fig. 5a provides a partial schematic interpretation of the two types of attention maps to facilitate understanding.

In Fig. 5b, the group–group self-attention assigns relatively higher weights to phenyl-related groups, while the group-to-molecule attention places greater emphasis on nitro groups. In Fig. 5c, the group–group self-attention highlights ether groups, whereas the group-to-molecule attention focuses more strongly on amide groups. The remaining cases (Fig. 5d–e) further illustrate distinct patterns of group–group and group–molecule interaction strengths across different mixtures, demonstrating that ChemFlow adaptively allocates attention according to the dominant chemical interaction motifs present in each system. Overall, these results demonstrate a clear division of labor among the three attention mechanisms: group–group attention models local functional group interactions, group-to-molecule attention injects salient group information into molecular representations. Together, they enable the model to learn chemically meaningful, multi-scale representations for complex mixtures.

**Limitations and Future Directions**

While ChemFlow offers a significant advance in the modeling of complex chemical systems, limitation remains. One notable challenge lies in the predefined group encodings and the construction of group-level hypergraphs based on chemical priors, which can introduce additional complexity in model construction. Moreover, this approach may overlook certain unique molecular structures that do not conform to the predefined group definitions ; incorporating adaptive group definitions could potentially address this issue. Furthermore, the inclusion of Neural Circuit Policies (NCP), while enhancing the model's ability to adapt atomic features, results in longer inference time compared to simpler architectures, e.g. the linear layers.

Looking ahead, pretraining ChemFlow on multiple, diverse chemical datasets could offer promising opportunities for the model to learn broader chemical knowledge. Such pretraining would allow ChemFlow to acquire rich, generalizable representations of chemical structures and interactions, which could then be fine-tuned to specific tasks, improving its adaptability and accuracy across a wider range of chemical systems. Additionally, integrating physical constraints directly into the training process via the loss function could further enhance the model's generalization capabilities, ensuring it is not only better suited for specific tasks but also more aligned with real-world chemical phenomena.

By incorporating these advancements, future improvement of ChemFlow could proceed toward even more powerful and versatile capability of handling increasingly complex and diverse chemical environments with greater efficiency and accuracy.

# Conclusion

ChemFlow offers a robust, multi-scale framework for modeling the intricate interactions that govern the behavior of chemical mixtures. By hierarchically linking atomic, group, and molecular representations, ChemFlow captures the core dynamics of both intramolecular and intermolecular interactions that dictate mixture properties. In comparative evaluations across a range of datasets, ChemFlow consistently achieves state-of-the-art performance, particularly in modeling concentration-sensitive behaviors. The model's ability to dynamically adjust embeddings at multiple levels—driven by the surrounding molecular environment and mixture concentration—ensures that it not only models group-level interactions within molecules but also how molecular interactions across mixtures influence each other. This hierarchical, concentration-aware feature flow mechanism is central to ChemFlow's capability to handle the complex, context-dependent nature of chemical systems. Crucially, ChemFlow goes beyond descriptive modeling, enabling the simulation of the fundamental interactions that underlie chemical behavior at multiple scales, positioning it as a powerful and generalizable tool for advancing predictive chemistry.

# Methods

**Graph Representation and feature of ChemFlow**

The following are detailed details about various features that were not mentioned in the main text:

To represent the true physicochemical states of atoms, functional groups, and molecules within complex chemical environments, we construct one molecular graph $\mathcal{G}_a^{(s)} = \left(\mathcal{V}^{(s)}, \mathcal{E}_{atom}^{(s)}\right)$ and two hypergraphs: (1) the group hypergraph $\mathcal{H}_{group}^{(s)} = \left(\mathcal{V}^{(s)}, \mathcal{E}_g^{(s)}\right)$ and (2) the molecular hypergraph $\mathcal{H}_{mol}^{(s)} = \left(\mathcal{V}^{(s)}, \mathcal{E}_{mol}^{s}\right)$.

For a mixture sample $s$, we consolidate atoms from all $K_s$ component molecules and concatenate them in a fixed order. Let $\mathcal{V}^{(s)} = \{v_1, ..., v_{N_s}\}$ denote the node set containing all $N_s$ atoms in the mixture, each node possesses features $h_i$. concatenated in a fixed order. The atomic graph is defined as:

$$\mathcal{G}_a^{(s)} = \left(\mathcal{V}^{(s)}, \mathcal{E}_{atom}^{(s)}\right),$$

where $\mathcal{E}_{atom}^{(s)}$ represents the set of intramolecular chemical bonds. For any bond $(i,j) \in \mathcal{E}_{atom}^{(s)}$, the edge feature is denoted as $e_{ij}$.

To capture local substructures, we model functional groups as hyperedges over the atomic node set $V^{(s)}$. We identify functional groups using SMARTS pattern matching. Let $G_s$ be the total number of matched group instances in sample $s$. For the $m$-th group instance ($m \in \{1, ..., G_s\}$), let $A_m \subseteq \{1, ..., N_s\}$ be the index set of its constituent atoms. The corresponding group hyperedge is defined as:

$$e_m^{group} = \{v_i \in \mathcal{V}^{(s)} \mid i \in A_m\}$$

The group hypergraph is then defined on the atomic node set as:

$$\mathcal{H}_{group}^{(s)} = \left(\mathcal{V}^{(s)}, \mathcal{E}_g^{(s)}\right), \text{ where } \mathcal{E}_g^{(s)} = \{e_m^{group}\}_{m=1}^{G_s}$$

The connectivity between atoms and groups can be formally represented by an incidence matrix $\mathbf{B}^{(s)} \in \{0,1\}^{N_s * G_s}$

$$B_{i,m}^{(s)} = \begin{cases} 1, & if\ v_i \in e_m^{group} \\ 0, & \text{otherwise} \end{cases}$$

*Implementation Note:* Since $\mathbf{B}^{(s)}$ is highly sparse, we store it in Coordinate Format (COO) as an edge list.

Analogous to the group level, we explicitly model each component molecule in the mixture as a hyperedge to learn global component-level representations.

Let the mixture contain $K_s$ component molecules. Since atoms are concatenated, each atom $v_i$ belongs to exactly one component molecule. We define the $k$-th molecular hyperedge as the set of atoms belonging to the $k$-th component:

$$m_k(i) \in \{0,1\}, \quad k \in \{1, ... ... K_s\}$$

The $k$-th molecular hyperedge is defined as

$$e_k^{mol} = \{v_i \in \mathcal{V}^{(s)} \mid \text{atom } i \text{ belongs to component } k\}$$

The molecular hypergraph is defined as:

$$\mathcal{H}_{mol}^{(s)} = (\mathcal{V}^{(s)}, \mathcal{E}_{mol}^s), \quad where \; \mathcal{E}_{mol}^s = \{e_k^{mol}\}_{k=1}^{K_s}$$

Equivalently, this relationship is represented by a molecular incidence matrix $\mathbf{S}^{(s)} \in \{0,1\}^{N_s * K_s}$:

$$S_{i,k}^{(s)} = \begin{cases} 1, if \; v_i \in e_k^{mol} \\ 0, \quad otherwise \end{cases}$$

*Implementation Note:* In practice, is implemented using boolean masks (e.g., mask_1, ..., mask_k), which effectively partition the node set $\mathcal{V}^{(s)}$ into disjoint components.

Additionally, to establish a hierarchy, we map each functional group to its parent molecule $k$, denoted as $\text{mol\_id}(m) = k$.

In the molecular graph, each atom is represented using three complementary descriptors:

(i) $h_{i-g}$: group assignment, specifying the functional group to which each atom belongs and capturing the hierarchical atom–group relationship;

(ii) $h_{i-b}$: environment descriptor, local chemical environment descriptor, summarizing bonded structural and electronic features that define the atom's immediate chemical surroundings; The atomic features ( $h_{i-b}$ ) considered are summarized as follows: (1) Symbol. (2) Number of Valence Electrons. (3) Number of Hydrogens. (4) Formal charge. (5) Hybridization. (6) Ring Membership (isInRing). (7) Aromaticity (isAromatic).

(iii) $h_{i-a}$: intrinsic physicochemical attributes, characterizing the elemental properties that underlie its baseline steric and electronic behavior. such as: (1) atomic number. (2) electronegativity. (3) covalent radius. (4) atomic mass. (5) first ionization energy. (6) electron affinity.

To encode functional-group information for both atoms and groups, we construct a custom SMARTS dictionary and perform substructure matching on each SMILES. The SMARTS dictionary is user-defined and specifies a vocabulary of functional fragments. In this work, we include a set of common groups such as phenyl rings, hydroxyl, carboxyl, carbonyl, ether, aromatic five-membered rings, amides, etc.; implementation details are provided in our code.

Given the SMARTS matches, we assign: (i) an atom-level group membership vector $h_{i-g}$ (one-hot or multi-hot when overlaps occur), (ii) a molecule-level global group encoding that counts the occurrences of each group type in the molecule (represented as a group-count vector), and (iii) group instances (matched substructures) used to construct the group hypergraph by treating each matched instance as a hyperedge over its constituent atoms.

Importantly, the SMARTS dictionary is not fixed: readers can adapt or extend the group vocabulary based on domain knowledge and the target properties of interest, yielding a more appropriate group partitioning and improving model suitability for specific chemical domains.

Group-level representations $G_m$ and molecular-level representations $M_k$ are obtained by aggregating atomic features, thereby preserving essential electronic and environmental information at higher structural levels.

**ChemFlow architecture**

**Chem-embed: Mixture-Induced Multimodal Atomic Representation.**

Firstly, we fuse the different features of each atom step by step, and the specific calculation process is as follows:

**Mixture → molecule:**

$$Q = \overrightarrow{H_{gi}} * W_Q,$$

$$K = \oplus_{\substack{j=1 \\ j \neq i}}^{N}(concat(\overrightarrow{H_{gj}}, C_{gj})) * W_K,$$

$$V = \oplus_{\substack{j=1 \\ j \neq i}}^{N}(concat(\overrightarrow{H_{gj}}, C_{gj})) * W_V$$

$$Attention = softmax\left(\frac{QK^T}{\sqrt{d_k}}\right) * V$$

$$\overrightarrow{H_{gi}} = LayerNorm(Q + Attention)$$

Here, $W_Q$, $W_K$ and $W_V$ denote the query, key, and value matrices used in computing the cross-attention mechanism, respectively; $C_{gj}$ represents the concentration of neighboring molecules.

**molecule → atomic group assignment:**

$$\overrightarrow{h_{\iota-g}} = LayerNorm(Q + softmax\left(\frac{QK^T}{\sqrt{d_k}}\right) * V)$$

where $Q = \overrightarrow{h_{\iota-g}} * W_Q$, $K = \overrightarrow{H_{g\iota}} * W_K$, $V = \overrightarrow{H_{g\iota}} * W_V$.

**Group assignment → atomic environment:**

$$\overrightarrow{h_{\iota-b}} = LayerNorm(Q + softmax\left(\frac{QK^T}{\sqrt{d_k}}\right) * V)$$

Where $Q = \overrightarrow{h_{\iota-b}} * W_Q$, $K = \overrightarrow{h_{\iota-g}} * W_K$, $V = \overrightarrow{h_{\iota-g}} * W_V$.

Atom-wise Feature Generation with NCP Units

We generate each atom's physical descriptors by integrating its environment-conditioned state and its local concentration.

The initial hidden state is: $\vec{H}^0 = cat(h_{i-b}, h_{i-g}, c)$

For NCP step k we compute: $\overrightarrow{h_{\iota-a}}^k, \vec{H}^k = NCP(\overrightarrow{h_{\iota-a}}^{k-1}, \vec{H}^{k-1})$

with the per-neuron update:

$$\overrightarrow{h_{\iota-a}}^k, \vec{H}^k = \sigma\left(f\left(\left[\overrightarrow{h_{\iota-a}}^{k-1}, \vec{H}^{k-1}\right]\right)\right) \odot \sigma\left(g\left(\left[\overrightarrow{h_{\iota-a}}^{k-1}, \vec{H}^{k-1}\right]\right)\right) + [1 - \sigma\left(f\left(\left[\overrightarrow{h_{\iota-a}}^{k-1}, \vec{H}^{k-1}\right]\right)\right)] \odot \sigma\left(h\left(\left[\overrightarrow{h_{\iota-a}}^{k-1}, \vec{H}^{k-1}\right]\right)\right)$$

in which $\sigma$ denotes the activation function, while $f$, $g$, and $h$ are learnable transformation functions.

The final atomic embedding concatenates multiple feature types:

$$\vec{h_\iota} = concat(\overrightarrow{h_{\iota-a}}, \overrightarrow{h_{\iota-b}}, \overrightarrow{h_{\iota-g}})$$

**Message-Passing Step of atom (performed at every iteration k):**

At each message-passing iteration, we first modulate atomic features with the global mixture context and then aggregate intra-molecular messages, concluding with a concentration-aware memory update.

(1) Feature-wise Linear Modulation: $\vec{h}_i^k = \vec{h}_i^k \odot \gamma(c) + \beta(c)$

where $\gamma(\cdot)$ and $\beta(\cdot)$ are learned parameters.

(2) Atom-level aggregation: $\vec{h}_i^k = \Theta \vec{h}_i^{k-1} + \frac{1}{N}\sum_{j \in N(i)} \vec{h}_j^{k-1} * (\sigma(W * \vec{e_{ij}}))$ which allows continuous update of atomic features upon integrating the information of neighboring atoms.

(3) Concentration-Aware Memory Update: $\vec{h}_i^k, \vec{H}^k = NCP(cat(\vec{h}_i^k, C), \vec{H}^{k-1})$

Note: The memory state $H(k)$ tracks concentration-aware evolution but does not feed into subsequent message-passing updates; it is used only at the end. The above concentration-aware is only used for predicting concentration dependent datasets and is not used for predicting non concentration dependent optical property datasets.

**Feature Aggregation and Interaction Computation**

**Atom → Group / Molecule aggregation with concentration modulation.** To aggregate atomic representations into functional-group and molecular representations, we adopt Set2Set as a permutation-invariant readout. For the $m$-th functional-group instance with member-atom set $A_m$ and the $k$-th molecule with atom set $M_k$, we compute:

To explicitly model composition effects, we apply concentration-conditioned feature modulation using learnable functions $\gamma(c)$ and $\beta(c)$:

$$G_m = \text{set2set}\{h_i \mid i \in A_m\}, \quad M_k = \text{set2set}\{h_i \mid i \in M_k\}$$

**Group self-attention (group–group interactions)**. We then perform self-attention over all functional-group embeddings $\{G_m\}_{m=1}^{G_s}$ to capture both intra- and inter-molecular group interactions:

$$G_m = LayerNorm(Q + softmax\left(\frac{QK^T}{\sqrt{d_k}}\right) * V)$$

Where $Q = G_m * W_Q$, $K = G_m * W_K$, $V = G_m * W_V$.

Bidirectional cross-attention (group ↔ molecule coupling). Next, we update groups and molecules via bidirectional cross-attention, ensuring that each group is influenced by all molecules in the mixture and each molecule is informed by all group-level semantics:

Molecule← Groups:

$$G_m = LayerNorm(Q + softmax\left(\frac{QK^T}{\sqrt{d_k}}\right) * V)$$

Where $Q = G_m * W_Q$, $K = M_k * W_K$, $V = M_k * W_V$.

Group← Molecule:

$$M_k = LayerNorm(Q + softmax\left(\frac{QK^T}{\sqrt{d_k}}\right) * V)$$

Where $Q = M_k * W_Q$, $K = G_m * W_K$, $V = G_m * W_V$.

Finally, we pool the updated molecule- and group-level representations into a mixture representation and then use a Multi-Layer Perceptron ($MLP$) model is employed to accomplish the final prediction, combining chemical and physical features to produce a robust estimation of physiochemical properties.

**Hyperparameter**

To evaluate the impact of various dataset modeling strategies and sampling methods, we conducted various tests on the activity coefficient dataset. As shown in Fig. 6a–6c, ChemFlow demonstrated fast convergence and high accuracy on the mixed dataset, suggesting that the combination of datasets with increasing complexity allowed the model to better capture composition-dependent behavior. In Fig. 6d, we compare the results of five-fold random cross-validation and five-fold stratified cross-validation using the activity coefficient datasets. Both methods yielded consistent MAE, RMSE, and $R^2$ values, confirming the robustness of ChemFlow under different cross-validation schemes.

We also investigated the effect of different hyperparameters on model performance. As shown in Fig. 6f, high learning rates caused convergence issues, while even a relatively low learning rate ensured stable model performance, highlighting ChemFlow's robustness. In Fig. 6g, the choice of batch size had minimal influence on the final prediction accuracy, suggesting that ChemFlow can achieve fast convergence without being highly sensitive to batch size, ensuring efficient training.

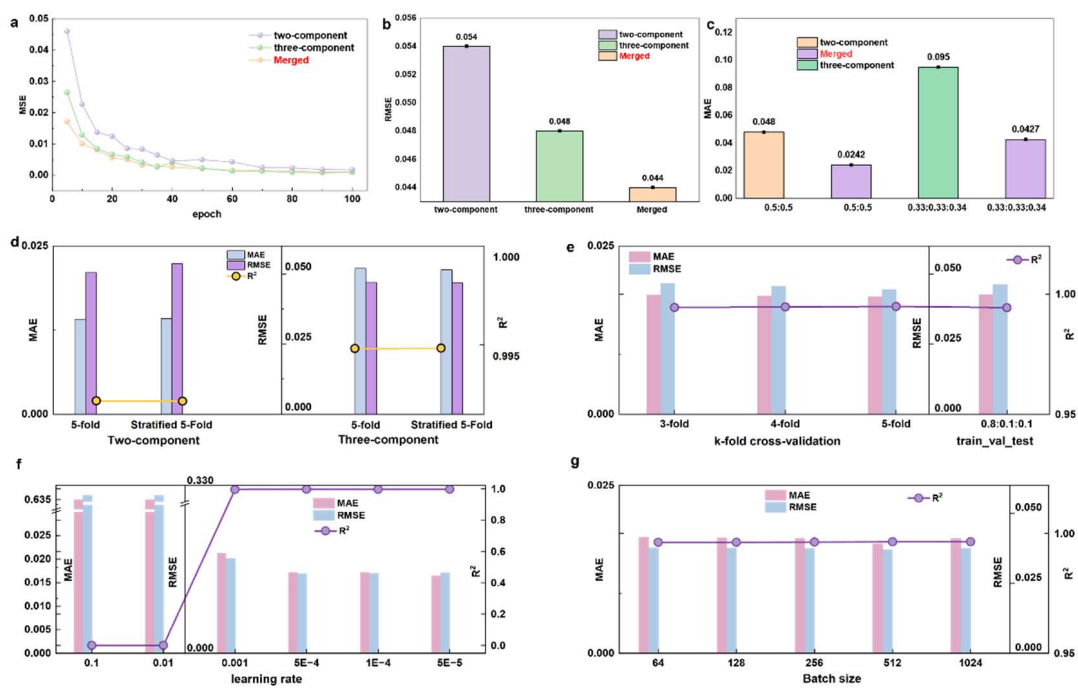

**Figure 6.** Influence of Dataset Modeling Strategies, Sampling Methods, and Hyperparameters. (a) The training curves, (b) error bars, (c) the extrapolation (to different concentrations) results, and (d) prediction results with different sampling strategies with different activity coefficient datasets; (e) prediction results with different dataset partitioning method; The influence of hyperparameters on the prediction results of mixed multi-component activity coefficients (f) learning rate; (g) batch size.

## Data and Code availability

Further information about the data and framework is also available at https://github.com/Fan1ing/ChemFlow.

## Acknowledgement

Generous financial support by the National Natural Science Foundation of China (U24A20527) and the Key Research and Development Program of Zhejiang Province (2023C01102, 2023C01208).